# Local gating of carbon nanotubes

M. J. Biercuk, N. Mason, C. M. Marcus,
*Department of Physics, Harvard University, Cambridge, MA 02138*

Local effects of multiple electrostatic gates placed beneath carbon nanotubes grown by chemical vapor deposition (CVD) are reported. Single-walled carbon nanotubes were grown by CVD from Fe catalyst islands across thin Mo "finger gates" (~150 nm × 10 nm). Prior to tube growth, several finger gates were patterned lithographically and subsequently coated with a patterned high-κ dielectric using low-temperature atomic layer deposition. Transport measurements demonstrate that local finger gates have a distinct effect from a global backgate.

Considerable effort has focused on incorporating single–walled carbon nanotubes (SWNTs) into nanoscale analogs of solid–state electronic devices. SWNT transistors have been realized [1,2,3], as have nanotube circuits exhibiting more subtle features such as Coulomb charging and the Kondo effect [4,5,6]. In order to fully explore the richness of nanotube device physics, independent control of relevant physical parameters is required. Many of these features may be controlled by electrostatic gating, in which the SWNT device is capacitively coupled to one or more nearby gate voltages. To date, however, independent parameter control via gating has not been realized; only *global* gating effects have been reported.

There have been a number of recent advances in gating of SWNT devices, including the use of Al backgates with thin oxide layers [7,8], the use of high-κ dielectrics [9], metallic side gates [10], liquid-phase electrolyte solutions [11], and external scanned gates [12,13,14]. However, a technique for implementing *local* gating via standard lithography with supporting transport data has not yet been presented to our knowledge. In previous work, nanotube devices with multiple electrostatic topgates [9] or a metallic gate underneath the nanotube [15] were fabricated to produce multigate devices, including OR logic transistors. In these cases, however, data appeared consistent with a global coupling of all topgates.

In this Letter we report local control of nanotube conduction via multiple electrostatic gates. Device fabrication is based on chemical vapor deposition (CVD) of SWNTs from Fe catalyst, and takes advantage of two notable processing features: (1) Thin Mo "finger gates" (~150 nm wide), robust against the CVD process, are defined lithographically, allowing nanotubes to be grown across them. (2) A high–κ dielectric layer is patterned by photolithography and a liftoff procedure using low-temperature atomic layer deposition (ALD) [16]. Transport data from a nanotube device fabricated in this manner indicate that the effect of individual finger gates is qualitatively different from that of a global backgate.

Devices were fabricated on doped Si wafers with 1 μm of thermally grown oxide as a base substrate, allowing the conducting Si to be used as a global backgate. Before nanotube growth, sets of five parallel Mo finger gates roughly ~150 nm wide and < 10 nm thick, spaced by ~400 nm, extending approximately 100 μm in length (Fig. 1), were patterned using electron-beam lithography liftoff and deposited using electron-beam evaporation. Larger Mo lines connected to the fine Mo gates were then patterned with photolithography liftoff.

Mo was chosen for its tolerance to the high temperatures and reducing atmosphere used in CVD processing, combined with reasonably low resistivity in thin-film form. Similar conclusions favoring Mo for this purpose were reached independently in Ref. [17]. Thin gate metallization (<10 nm thickness) was used to avoid bending defects created by a nanotube "draping" over raised contacts [18]. We found that 5 nm films of Mo exposed to CVD processing vanished, while thicker layers remained intact (minus ~5 nm). Thus, metal which was exposed to the CVD environment always included a ~5 nm sacrificial layer.

After fabrication, the finger gates and their connections were covered by 25 nm of $HfO_2$, deposited using low–temperature ALD and patterned using photolithography and liftoff [16]. The dielectric layer was patterned to form large mesas that covered the finger gates but left the contacts exposed, as shown in Fig. 2. Next, rectangular patterns (~1 µm×5 µm) were defined in a poly(methyl methacrylate) (PMMA) layer using electron-beam lithography, and ~1 nm Fe was deposited using thermal evaporation. The rectangles were oriented in rows on either side of the Mo finger gates, and served to locate the Fe catalyst to promote nanotube growth across the underlying finger gates. A standard CVD recipe using methane as a carbon source was employed for tube growth [19], after which SWNTs crossing the finger gates were located using an atomic force microscope (AFM) [20]. Finally, SWNTs were contacted with Ti/Au contact pads to complete the devices (Figs. 1, 3). Typical device dimensions (between contacts) were 3 – 5µm. Atomic force and (post–measurement) scanning electron microcopy ensured that the finger gates were continuous.

Transport measurements were made at 4K using a dc voltage bias, V = 10 mV, and measuring dc current, I. Data is presented for a single device (Fig. 3); similar behavior was observed for other devices. Conductance, G = I/V, was measured as a function of voltages applied to various finger gates and backgate. Sweeping the voltage on individual finger gates produces smooth changes in G (Fig. 4a). Most finger gates exhibit field–effect behavior, (F1, F3, F4), while one of the gates (F5) exhibits a broad resonance feature. Gate F1, located under the SWNT-metal contact, is likely tuning the transparency of the Schottky barrier (Fig. 4a). The general tendency for G to decrease as gates become more negative indicates that the nanotube doping is n-type.

Sweeping the backgate with the finger gate voltages held fixed produces a qualitatively different behavior in conductance. In this case, rapidly varying, reproducible fluctuations in G are found as a function of backgate voltage, $V_B$ (Fig. 4b). Setting a single finger gate to a nonzero voltage, $V_F$, with the other finger gates held at zero yields similar rapid fluctuations in $G(V_B)$, but with different overall amplitude, consistent with the $G(V_F)$ from Fig. 4a acting as an overall smooth envelope of $G(V_B)$. Examples of $G(V_B)$ for two settings of $V_F$ on F4 are shown in Fig. 4b; similar behavior was observed with other finger gates. The rapid fluctuations in $G(V_B)$ are presumably due to Coulomb blockade resulting from quantum dots defined by defects along the tube. The qualitative difference between the effects of the back gate and finger gates suggests that the finger gates act to locally tune the transparency of scattering centers in the SWNT while the back gate alters the electron configuration on the multiple dots [14]. This picture is supported particularly by the nonmonotonic (resonant-like) behavior of $G(V_{F5})$. Local scatterers have previously been linked to the formation of intratube quantum dots [14, 21, 22] and have been observed by scanned gate measurements [12, 13, 14] and electrical-force microscopy [23]. The absence of rapid Coulomb charging fluctuations in $G(V_F)$ suggests a model where each finger gate acts locally, tuning the transparency of a single defect along the tube. If the finger gates were instead having a global effect and coupling to the entire tube device, one would expect Coulomb-blockade phenomena very similar to those caused by sweeping the backgate, though perhaps on a different overall voltage scale.

Figure 4c shows device conductance as a function of both backgate and finger gate voltages for the case where all finger gates are swept together. Fluctuations in $G(V_B)$ with $V_F = 0$ V previously described appear again but now evolve continuously into oscillations in $G(V_F)$ with $V_B = 0$ V, demonstrating the approximately additive behavior $V_B$ and $V_F$ when all finger gates are swept. Evidently, when all finger gates are swept, they together do produce an effective global gating effect much like the backgate, albeit on a reduced voltage scale (as expected given the distances and dielectric constants). Thus although the effect of the individual finger gates is spatially localized along the nanotube, the area of influence is larger than that defined by the physical dimensions of the finger gates.

As a direct comparison, Fig. 4d shows corresponding plots when sweeping just one of the finger gate with the other finger gates held at 0 V. In this case, there is no additive effect between finger gate and back gate, even over an expanded range of $V_F$. Horizontal slices of the 2D plot show roughly the same behavior in $G(V_F)$ as observed at $V_B = 0$ V in Fig. 4a (ignoring switching noise) while vertical slices show that oscillations in $G(V_B)$ persist for all values of $V_F$.

In summary, we have demonstrated local gating using finger gates beneath a catalyst-grown single-wall nanotube. The fabrication process takes advantage of robust Mo finger gates and liftoff-patterned dielectric films deposited by low-temperature atomic layer deposition. Future applications of the technique reported include fabricating multigate nanotube FETs or quantum dots.

(After this work was completed, related results were reported in Ref. [24]. Ref. [24] focuses on local gating in nanotube field effect transistors using top gates rather than undergates. In contrast to our devices, the devices in this case were created using random nanotube deposition out of solution. This technique does not allow for much control over the physical location of the device, an important element in the incorporation of nanotube electronics into more advanced circuits.)

We wish to thank J. S. Becker for her help with ALD and D. J. Monsma for many helpful discussions. This work was supported by funding from the NSF through the Harvard MRSEC and EIA-0210736, and the Army Research Office, under DAAD19-02-1-0039 and DAAD19-02-1-0191. M.J.B acknowledges support from an NSF Graduate Research Fellowship and from an ARO Quantum Computing Graduate Research Fellowship. N.M. acknowledges support from the Harvard Society of Fellows.

Figure Captions:

**Fig.1**.  Schematic of finger gated devices.  Mo gates (150 nm wide × 10 nm thick) were defined lithographically on a Si/SiO$_2$ substrate and subsequently coated with 25 nm of HfO$_2$ grown by low-temperature ALD.  Nanotubes were grown across these local gates by CVD and contacted with Ti/Au electrodes.  Not to scale.

**Fig. 2**.  a) Scanning electron micrograph showing complete device including Al wire bonds.  Note liftoff-patterned ALD oxide mesa. b)  Higher magnification micrograph of ALD mesa edge (middle) showing Ti/Au wires on top of the mesa (upper left) and Mo wires running underneath the patterned ALD (bottom).

**Fig. 3a/b**.  Atomic force micrographs of nanotubes grown across Mo finger gates and contacted (far left and far right) by Ti/Au leads. Note one finger gate passes directly underneath the nanotube-metal contact.  Arrows indicate the location of the nanotube

**Fig. 4**. Transport measurements taken from the device depicted in Fig. 3. All data taken at 4K. a) Conductance as a function of various finger gate voltages.  Each trace represents the effect of a single finger gate swept from +4 V to –4 V while all others, including the backgate, are set to 0 V.  Gate F2 showed significant leakage above $V_{F2}$ ~2 V and so was not included in these plots.  b) Charging effects observed by sweeping the Si backgate. Traces are displayed for two different voltages on finger gate F4, which changes the overall magnitude of the rapid fluctuations without changing the qualitative structure. c) Color plot of conductance as a function of backgate voltage ($V_B$) and common finger gate voltage ($V_F$) (i.e. all finger gates swept together) indicating an additive effect of $V_B$ and $V_F$.  Color scale shows conductance in units of $e^2/h$. d)  Comparable color plot showing conductance as a function of $V_B$ and a *single* finger gate at $V_F$ with other finger gates set to V=0.

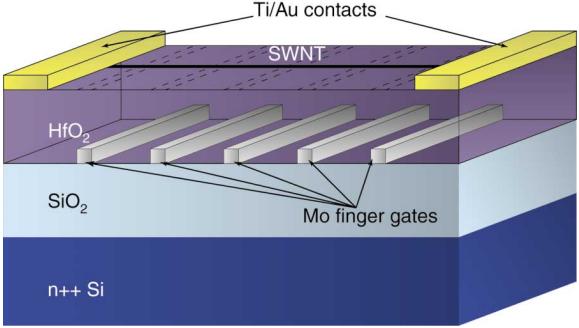

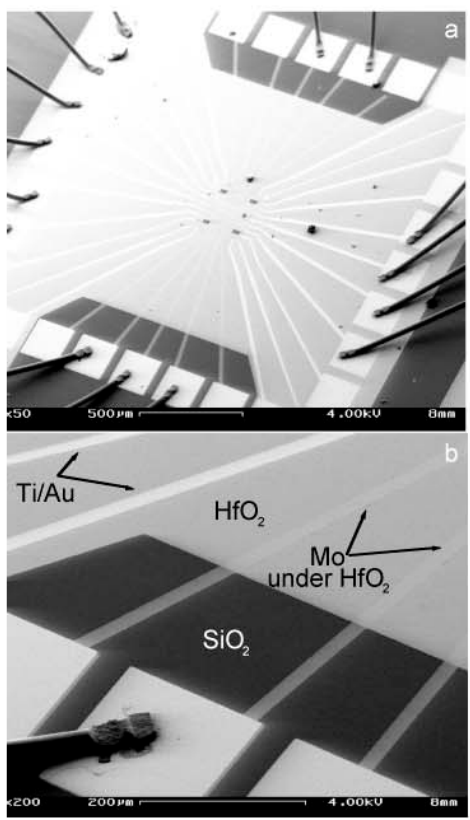

M.J. Biercuk et. al.
  Figure 3.

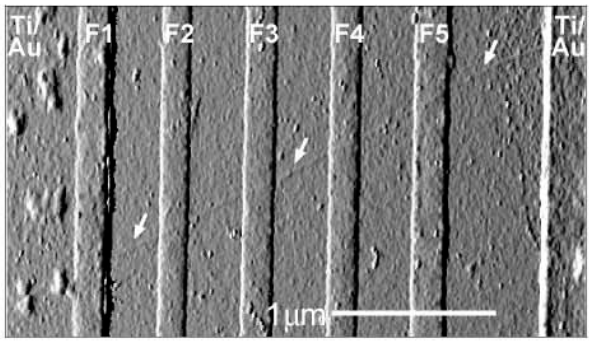

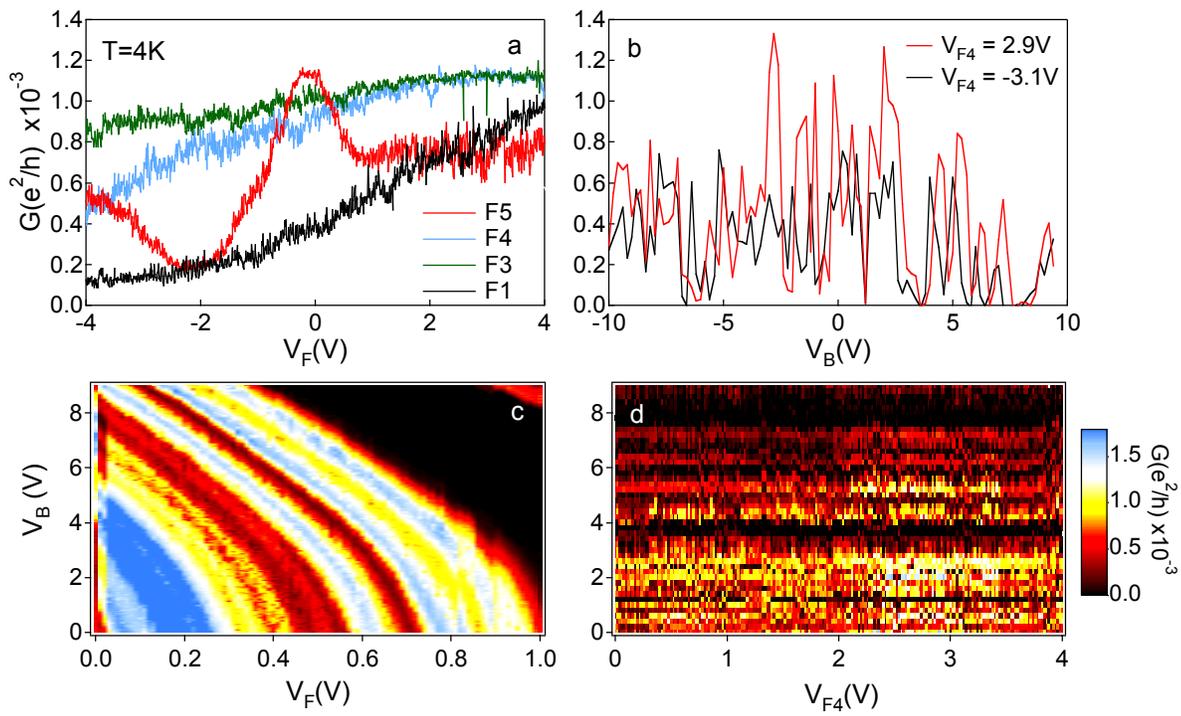

M. J. Biercuk et. al.
Figure 4.